\documentclass[conference,10pt]{IEEEtran}
\usepackage[utf8]{inputenc}
\usepackage[T1]{fontenc}
\usepackage[scaled = .88]{helvet}
\usepackage[caption=false,font=footnotesize]{subfig}
\usepackage{
  booktabs,
  graphicx,
  kantlipsum,
  microtype,
  multirow,
  url,
  xcolor,
  xspace,
  tcolorbox,
  tabularx
}
\usepackage{framed}
\usepackage{algorithmic}
\usepackage{multirow}
\usepackage{balance}
\usepackage[normalem]{ulem}
\useunder{\uline}{\ul}{}
\usepackage[capitalize]{cleveref}
\newboolean{showcomments}

\setboolean{showcomments}{true}

\ifthenelse{\boolean{showcomments}}
  {\newcommand{\nb}[2]{
    [#1: \emph{#2}]
   }
  }
  {\newcommand{\nb}[2]{}
  }

\newcounter{rqcounter}

\newcommand{\researchq}[2]{
  \smallskip
  \noindent
  \refstepcounter{rqcounter}\label{rq:#1}%
  \textit{RQ\,\therqcounter:} \textit{#2}
  \medskip
}

\newcommand{\uvic}{University of Victoria} 
\newcommand{\refrq}[1]{\emph{RQ \ref{#1}}\xspace}

\newcommand{\etal}{{\emph{et al.}}\xspace}
\newcommand{\eg}{{\emph{e.g.},}\xspace}
\newcommand{\ie}{{\emph{i.e.},}\xspace}

\newcommand{\variable}[1]{\textsf{#1}}

\newcounter{implcounter}

\definecolor{blueish}{cmyk}{61, 47, 0, 40}

\begin{document}
\title{Autonomy Is An Acquired Taste: Exploring Developer Preferences for GitHub Bots}

 \author{
\IEEEauthorblockN{Amir Ghorbani\IEEEauthorrefmark{1},
Nathan Cassee\IEEEauthorrefmark{2},
Derek Robinson\IEEEauthorrefmark{1},
Adam Alami\IEEEauthorrefmark{3},\\
Neil A. Ernst\IEEEauthorrefmark{1},
Alexander Serebrenik\IEEEauthorrefmark{2},
Andrzej Wąsowski\IEEEauthorrefmark{4}}
\IEEEauthorblockA{\IEEEauthorrefmark{1} University of Victoria, Canada \{amirrezaghorbani, drobinson, nernst\}@uvic.ca}
\IEEEauthorblockA{\IEEEauthorrefmark{2} Eindhoven University of Technology, The Netherlands \{n.w.cassee, a.serebrenik\}@tue.nl}
\IEEEauthorblockA{\IEEEauthorrefmark{3} Aalborg University, Denmark adal@cs.aau.dk}
\IEEEauthorblockA{\IEEEauthorrefmark{4} IT University of Copenhagen, Denmark wasowski@itu.dk}
}
\maketitle

\begin{abstract}

Software bots fulfill an important role in collective software development, and their adoption by developers promises increased productivity. 
Past research has identified that bots that communicate too often can irritate developers, which affects the utility of the bot.
However, it is not clear what other properties of human-bot collaboration affect developers' preferences, or what impact these properties might have.
The main idea of this paper is to explore characteristics affecting developer preferences for interactions between humans and bots, in the context of GitHub pull requests. 
We carried out an exploratory sequential study with interviews and a subsequent vignette-based survey. We find developers generally prefer bots that are personable but show little autonomy, however, more experienced developers tend to prefer more autonomous bots.
Based on this empirical evidence, we recommend bot developers increase configuration options for bots so that individual developers and projects can configure bots to best align with their own preferences and project cultures.

\end{abstract}

\begin{IEEEkeywords}
Software Bot, Pull Request, Human Aspects%
\end{IEEEkeywords}

\section{Introduction}

\noindent
Software bots can increase developer productivity\,\cite{peng2019exploring, wessel2020expect}. Perhaps this is why bots are on the rise in collective software development\,\cite{Santhanam2022}. Bots are used as front-ends for static analysis tools, for security screening, and for development process management, for instance closing stale issues. However, since software bots are relatively new, we do not yet understand the behavioral patterns that make developers accept or reject working with bots. Existing research focuses on understanding specific bots~\cite{mirhosseini2017automated}, characterizing bot types~\cite{storey2016disrupting, wessel2018power} and bot actions~\cite{peng2019exploring,wessel2020expect} rather than on the human--bot collaboration process.
While it is known that bots irritate humans with too frequent notifications~\cite{wessel2020inconvenient}, it is not clear what other features of bot communication matter for developers.

In this work, we explore \textbf{how specific bot behaviors affect developer perception of bots}, with the goal of \textbf{supporting informed designs for well-behaved software bots}. We begin with a broadly scoped exploratory interview study (\cref{sec:phase1}) addressing the following research question:

\researchq{interview}{What bot characteristics shape human perceptions of bot behavior?}

\noindent
Our interviews identified the bots our respondents are familiar with.
We then probed the aspects of bot behavior the respondents preferred.
Respondents indicated that even for bots performing highly useful tasks, certain aspects of the bot's behavior affected developer perceptions. Our interviews led to a set of themes that highlight several important factors, including the bot role (task), the degree to which a bot acts autonomously (\emph{Autonomy}), and the \emph{Persona} presented by a bot.
As bot roles (Task) have been well-studied by Erlenhov \cite{erlenhov2020empirical}, Wessel \cite{wessel2018power}, and others, we focused the second stage of this work on Autonomy and Persona.
Following an exploratory sequential design \cite{creswell2017designing}, we formulated the second research question:
\looseness -1

\researchq{survey}{How does the degree of Autonomy (resp. Persona) influence a developer's Preference for the bot's actions in a pull request discussion?}

\noindent
For this question, we conducted a survey using a custom vignette-based instrument to evaluate the influences of the two independent variables, persona and autonomy, on user preferences (\cref{sec:phase2}). Within the survey each vignette presented the respondents with two nearly identical scenarios shown side-by-side. The two scenarios showed respondents a bot interacting on a pull request discussion in a software project on GitHub. Critically, we varied the bot's behavior to contrast aspects of either autonomy or persona so that each side of the vignette represented a different pole of the construct. We then asked developers which of the two sides they preferred.

\medskip

\noindent
The key contributions of this study include:
\begin{itemize}

  \item A set of themes, derived from practitioner interviews, identifying key aspects influencing perceptions of human-bot interactions.

  \item The identification and definition of the constructs of bot Autonomy and Persona, along with a specific operationalization of those constructs.

  \item Evaluation of the effect of Autonomy and Persona, using a custom-designed vignette-based instrument allowing to increase realism and control for studies of bot behavior.

  \item A set of statistically validated models with quantitative evidence for the preferences of developers for specific bot behavior. Allowing us to conclude that developers generally prefer reactive, personable bots.

  \item Actionable recommendations grounded in our empirical evidence. Our recommendations set future directions on the use of bot on software development activities. We recommend that the the \emph{autonomy} and \emph{persona} of bots should be fluid and left to developers to align according to their preferences and tasks at hand.

\end{itemize}

\noindent
Our findings open new avenues for research into how to design bot personas for particular projects and how to tailor them for different demographics of developers. They suggest a focus on acceptable autonomy.%
\looseness = -1

\section{Background and Related Work}

\noindent
Storey and Zagalsky\,\cite{storey2016disrupting} define a software bot ``as a conduit or an interface between users and services, typically through a conversational user interface (UI).''
Notions of interaction and intelligence are a dominant theme in distinguishing bots from other development tools~\cite{erlenhov2020empirical, wyrich2021waiting}.
In fact, Erlenhov \emph{et al.} found that there are three different definitions used by developers to define bots~\cite{erlenhov2020empirical}, based on whether developers perceived their utility as primarily for Chat, to perform Smart actions, or how they acted Autonomously.

Bots are widely adopted and seen as useful: both Wessel \etal and Peng and Ma have found that the adoption of bots by open-source software projects results in more pull-requests being merged, leading to a more efficient division of work~\cite{wessel2020expect,peng2019exploring}.
However, the introduction of bots is not without risks, as Wessel \etal found that
developers describe noise as a central and recurrent problem when describing bots~\cite{wessel2021dont}.
To counteract this, Wessel \etal propose the usage of an automated moderator that filters and aggregates bot actions~\cite{wessel2022bots}.
While the work of Wessel \etal focuses on the annoying behavior of bots on social coding platforms, we are more specific, studying the specific bot constructs that developers prefer bots exhibit.

Autonomy and persona are the two characteristics that emerged from the interview study and explored further in the experimental survey.%

\noindent\textbf{Autonomy.} 
While some people may want highly autonomous software bots, they are in the minority of software bot users \cite{erlenhov2020empirical}.
Seiffer \emph{et al.} found that bots who are too autonomous provoke skepticism from their users, as the users cannot trace and manage the bots' work \cite{seiffer2021understanding}. 
This skepticism was also found by Liao \etal outside of software engineering, where professionals who
experienced an automated personal assistant were averse to a more proactive agent because of the risk of interruptions~\cite{liao2016personalai}.
When it comes to the evaluation of automated assistants, Schaffer \etal found that self-reported experience of professionals influences how willing participants were to accept
assistance from automated agents~\cite{schaffer2019icandobetter}. As there are examples of highly proactive bots in software engineering (\eg Dependabot) we aim to understand whether developers prefer reactive or proactive bots, as 
most existing literature studies notions of autonomy outside software engineering.
Meanwhile, both the work of Erlenhov \etal~\cite{erlenhov2020empirical} and Seiffer \etal~\cite{seiffer2021understanding} does not evaluate the preference of developers in an experimental set-up. 

\noindent\textbf{Persona}. 
While many software bots communicate through text-based conversational UI, they do not have to take on science fiction robots' prototypical dry and analytical persona as Nass \etal found that
humans respond to computer generated cues with social behavior~\cite{nass1994casa}.
Farah \emph{et al.} found that software bots which harness humour in the form of puns appear friendlier and more full of personality \cite{farah2021conveying}.
Furthermore, software bots which display higher levels of anthropomorphism build more trust with their human counterparts by creating a sense of familiarity. This has been found both within software engineering ~\cite{seiffer2021understanding, li2021machinelike, pfeuffer2019anthropomorphic} and outside of software engineering~\cite{jain2018evaluatingchatbots, chaves2020chatbot, chaves2022chatbotdesign}.
However, it might not always be needed for bots to have higher levels of anthropomorphism,
as  Clark \etal found that humans still approach
bots and interactions with bots as fundamentally different from human interactions~\cite{clark2019goodconvo}.
While we know that anthropomorphic bots build more trust and appear friendlier it has not yet been studied whether developers prefer bots that use more personable language. Especially not in an experimental set-up mimicking the 
interface of GitHub where developers are shown a direct comparison between a personable and factual bot and asked to indicate a preference.

\section{Phase I: Perceptions of Bot Behavior}
\label{sec:phase1}

\begin{table}[t]
\centering
\caption{Interview Participants Demographics}
\label{tbl:participant}
    \begin{tabular}{lllll}
    \toprule
    Community & ID  &  Role & Experience & Country \\
    \midrule
    \multirow{7}{*}{FOSSASIA} & P1  & Maintainer & 4 Years  & India      \\
    & P2  & Newcomer & 3 Years & India      \\
    & P3  & Contributor  & 6 Years & India      \\
    & P4  & Maintainer & 1.5 Year & India      \\
    & P5  & Contributor & 5 years & Germany    \\
    & P6  & Maintainer & 6 Years & Singapore  \\
    & P7  & Maintainer & 4 Years & Sri Lanka  \\ \hline

    \multirow{1}{*}{ROS} & P8  & Contributor & 12 years & USA        \\ \hline

    \multirow{3}{*}{Coala}  &  P9 & Contributor       & 4 Years & India      \\
    & P10  & Contributor & 2 Years & India      \\
    & P11  & Contributor  & 5 years & Japan      \\ \hline

    \multirow{1}{*}{RTEMS Community} &  P12  & Contributor      & 3 Years & India      \\ \bottomrule
    \end{tabular}
\end{table}

\noindent
We implemented an exploratory sequential design \cite{creswell2017designing} to answer our two research questions. The qualitative phase (interviews) produced themes investigated in the quantitative phase (Sect. \ref{sec:phase2}, surveys). In Phase I, described in this section,  interviews produced preliminary findings of human perceptions of bots' characteristics and behavior.

Phase I findings guide the design of the second phase (Section \ref{sec:phase2}). This design ({Qualitative} $\rightarrow$ {Quantitative}) allowed us to draw grounded, reliable findings in Phase I. In the second phase we used survey instruments and a larger sample to broaden the empirical coverage, control for extraneous variables, and test some of the earlier claims.
Since there are no clear theories about software bot behavior and developer perceptions, we position the goals of the study as exploratory.

\subsection{Interviewing Developers}

\noindent
We interviewed twelve open source software developers from four communities in order to understand their perception of bot behavior in open source communities and address \refrq{rq:interview}.

\subsubsection*{Sampling \& Recruitment} We recruited participants for our interview research via convenience sampling \cite{patton2014qualitative}. We asked three FOSSASIA, ROS, and Coala maintainers to assist us attract contributors and maintainers from respective communities. We have long-standing relationships in the communities we chose, and we sought out to our contacts to assist us recruit participants. We prepared an invitation to participate, which was then sent to a group of contributors and maintainers through two of our connections (FOSSASIA and Coala). Instead, our ROS contact suggested possible candidates.

We contacted four possible ROS participants, and our connections in FOSSASIA and Coala issued twelve and ten invitations on our behalf to community contributors and maintainers, respectively. We successfully interviewed one ROS participant, seven FOSSASIA participants, and three Coala participants. One individual is active in both the Coala and RTEMS communities, but chooses to identify with the latter.

\subsubsection*{Participant Demographics} Table \ref{tbl:participant} highlights the demographics of our interviewees. The first column is the community to which the participants contribute. The third column is their roles in their respective communities (either contributors or maintainers) and ``Experience'' is the accumulated number of years they have been contributing to their communities. ``Country'' is the country of residence.

\begin{table}[t!]
\footnotesize
\caption{Most salient interview questions. Q1 and Q2 were brief introductory questions, not shown here.}%
\renewcommand\arraystretch{1.4}
\label{tbl:interviewguide}
  \begin{tabular}{p{8cm}}
    \toprule
    \textbf{Core questions}\\
    \midrule
    Q3 - Are you using bots in your community? Have you used bots before? For what purpose? How?\\
    Q4 - Does it make a difference for you to deal with a Bot or a human in the context of contributing to your community and how? \\
    Q5 - How would you feel if a bot immediately rejected your PR? \\
    Q6 - How would you feel if a bot immediately accepted your PR? \\
    Q7 - How would you feel when a Bot submit a PR?\\
    Q8 - How would you expect from a Bot to behave, concretely?\\
    Q9 -  What opportunities do you see for bots in your project/community? \\
    \midrule
    \textbf{Examples of probing questions}    \\
    \midrule
    Q10 - Do you have an example [of bots in your community] you can share with us? \\
    Q11 - Would your attitude change in the review process [if a bot was to accept the PR], for example? \\
    Q12 -  Have you had a similar situation [where a bot rejected/accepted a PR]? Can you share it with us? \\
    Q13 - What should the bot do to not annoy you? \\

    \bottomrule
    \end{tabular}
\end{table}

\subsubsection*{Data Collection}
To balance the need to obtain rich data and to maintain focus during the interview we opted for semi-structured interviews to collect Phase I data.
We used a predefined interview guide to semi-structure the interviews (\cref{tbl:interviewguide}), and prepared probing questions (e.g., ``not annoy you'') to trigger the participants to share further detailed accounts of their experiences.

The aim of the interview study is to collect data based on contributors' and maintainers' experiences using bots in their respective communities' pull request (PR) process. This is a relevant source of data and the approach allowed access to an \emph{emic} perspective, points of view representing the meaning people give to events, relationships, behaviors, and experiences. Data collected through semi-structured interviews provide insider information or knowledge about what happens in practice and how people perceive the events around them, which is difficult to obtain otherwise~\cite{vanover2021analyzing}. 

Table \ref{tbl:interviewguide} documents key questions of the interviews.
To avoid socially desirable (i.e., providing responses that researchers like to hear) \cite{furnham1986response} and abstract responses, we planned questions seeking granular and detailed answers (e.g., Q5--Q8) prompting the interviewees to share relevant accounts from their experiences. The  detailed interview guide is available in the replication package (Sect. \ref{sec:replication}).
We used Zoom, a virtual meeting tool, to conduct the interviews.
Upon the completion of an interview, we transcribed the recording using Temi,\footnote{\url{https://www.temi.com}} an online transcription tool.
We manually checked the recording against the transcripts when the verbatim was unclear. The interviews lasted on average 60 mins and the transcripts average 20 pages. We obtained permission to make anonymized version of the interviews available as part of our replication package. The interviews were conducted between October and December 2020.
\looseness = -1

\subsection{Coding Transcripts}\label{sec:coding}

\noindent
Content analysis is the process of categorizing verbal or behavioral data to classify, summarize and tabulate the data. We opted for an inductive analysis approach \cite{miles1984qualitative}.
It is often used when there is limited understanding of the research phenomenon. Inductive analysis aims at generating meanings from the data collected in order to identify patterns and relationships to build a theory. Our study used a two-step analytical process as per recommendation~\cite{miles1984qualitative}.

\subsubsection*{Step 1: Developing Codes}
In this initial phase of the coding, we coded the data line-by-line, and we read the interviews text interpretively using \refrq{rq:interview} as an analytical lens. In parallel, we assigned meaningful codes to segments of the text. The outcome of this first step of coding is a codebook intended as an input to the subsequent phase of coding. We used ATLAS.ti\footnote{\url{https://atlasti.com}} to ingest the transcripts and manage the codebook. The full code list is available in our replication package (Sect. \ref{sec:replication}).

\begin{table*}[t!]
    \footnotesize
    \renewcommand\arraystretch{1.3}
    \label{tbl:codegroups}
        \begin{tabularx}{\textwidth}{p{0.27\columnwidth}p{0.53\columnwidth}X}
        \toprule
        Theme (freq.) & Definition & Evidence from the data\\
        \midrule
        \emph{Attitude}   (104)  &  How a human perceives bot actions  & \emph{``So, it's totally fine with me if the [bot] reviews and the bots review and say, you know, this is not fine''} (P10). \\

        \emph{Autonomy} (41)   & Does a human have to invoke/moderate a bot   & \emph{``... if [a bot] just goes and tells you something's broken, it's not as good as if it can, like, fix it for you''} (P8).\\
        & & \emph{``a bot is only as good as the automation it gives you''} (P8). \\

        \emph{Persona} (67)     & Aspects of a bot's character as perceived by a human  & \emph{``It [a bot] should be a little cool. Some sarcasm, some funny joke so to keep up the energy and enthusiasm in development''} (P3). \\

        \emph{Task}   (144)      & Tasks which a bot can perform          & \emph{``the bot has been configured to automate the various tests. So, let's say you create a pull request, then what happens that depending on the dependencies, and the CI/CD workflows of the of your particular project, whatever pull request someone is making,...And they might be using some custom tests like Travis CI, so that is something that these custom bots do it can run some tests that are being built particularly for that particular organization''} (P1).                       \\ %

        \emph{Feelings}  (28)   & Emotional response of a human evoked by either a bot or a human  & \emph{``\ldots in case if it is a bot [reviewing the PR] \ldots and he's [sic] simply rejecting my PR, then I will feel a little bad, and my perception towards that bot will actually change.''} (P3). \\

        \emph{Project Norm} (32) & Implicit practices within an OSS project        & \emph{``RTEMS is actually a community that works over emails, as of now we don't use bots''} (P12)              \\
        \
        \emph{Role}   (14)      & The role(s) a bot or human play within a project      & \emph{`So, this bot can do [style checks]. It is time-consuming to ask the reviewer for his review \ldots We can say the bot becomes one of the reviewers. ''} (P2).         \\
        \bottomrule
        \end{tabularx}

        \medskip
        
        \caption{Themes, their Definitions, and evidence from the data}%
    
    \end{table*}

\subsubsection*{Step 2: Categorization of Codes}
We identified patterns among the codes identified in the earlier phase of coding, and categorized them into themes.
Using ATLAS.ti, two authors of this paper performed the inductive coding approach described above. In total, they conducted six rounds of developing codes (Step 1). In the first three rounds, one interview transcript was coded by each author with a discussion session performed afterwards. In round four, no interview transcripts were coded, however, the authors merged codes together and revised the code list. In the fifth round, one author finished coding their remaining interviews, with the other author doing the same in the sixth round. After the sixth round, both authors categorized the codes together to form the final code list.

Since during each round of this coding process debriefing sessions were organized between the coders to discuss the disagreements, resolve them, and come to a consensus on the low-level codes, IRR measures are not applicable to this constructivist and emergent process \cite{empiricalstandards}.

\subsection{Findings From Phase I}

\noindent
The interviews provided a rich set of insights into how software developers perceive software bots. Seven themes emerged: \emph{attitude}, \emph{autonomy}, \emph{persona}, \emph{task}, \emph{feelings}, \emph{project norm}, and \emph{role}. Table~\ref{tbl:codegroups} defines these themes.

The respondents identified that bots play significant \emph{roles} in their projects, being seen primarily as assistants. Bots have a big influence on community perceptions and \emph{project norms}, leading people to \emph{feel} different emotions, whether annoyed, discouraged, or happy, among others. When we look at the \emph{attitudes} behind these feelings, the biggest were how independent the bot was, and how it came across. Both \emph{autonomy} and \emph{persona} affected the interviewees \emph{feelings} and \emph{attitudes} towards a given bot.

For example, \emph{``I've seen a lot of bots being very impersonal. Like bots are programmed to be impersonal. I personally don't like that''} (P6).
Bots need to be accepted by the project community in order to be effective, regardless of what they can accomplish: \emph{``even though [the bot is] rejecting my PR, that means he is able to understand what is a problem in my PR, but it is not intelligent enough to write that problem in the comment section and wait for the entire development''} (P3).

In Phase 1 data, the themes of \textbf{autonomy} and \textbf{persona} seem to evoke the strongest reactions amongst our interviewees.
For example, the lack of rational and factual explanations subsequent to bot actions can have ramifications on how the developer feels towards the bot's actions.
P3 expressed disappointment if a bot would reject his PR with no justification for such action. They stated: \emph{``in case if it is a bot doing the task, and a bot is not commenting on what error I made, and he's simply rejecting my PR, then I will feel a little bad, and my perception towards that bot will actually change''} (P3). P9 expressed stronger action: \emph{
``I think as a newcomer, it feels a lot worse, because if it's my first contribution and a bot rejected me, I'm not contributing more to this project. I'm done with this project} (P9).

Similarly, too much autonomy takes joy out of the process of contributing. For example, P2 and P6 explained that they would not feel the same ``joy'' if their PRs were accepted by a bot. They explained: \emph{``so, the thing with PRs is that you get the joy of finally getting merged after a lot of suggestions from the reviewers. And that joy is really not comparable for a bot just merging it with any reviews."} (P2) and \emph{``I might have probably been more happier [sic] if a human being accepted my PR''} (P6).

We concluded that \emph{autonomy} and \emph{persona} exert more influence in shaping developers perception of bots.
Hence, in phase 2, we decided to focus on these two variables, leading us to formulate \refrq{rq:survey}, to further our understanding of the influence of autonomy and persona on developer perception and expectation of bots in their pull request process.

\section{Phase II: Testing Bot Perceptions}\label{sec:phase2}

\begin{figure}[t!]

  \centering
  \includegraphics [
     width = \linewidth,
     clip,
     trim = 1.9mm 0mm 0mm 1mm
  ] {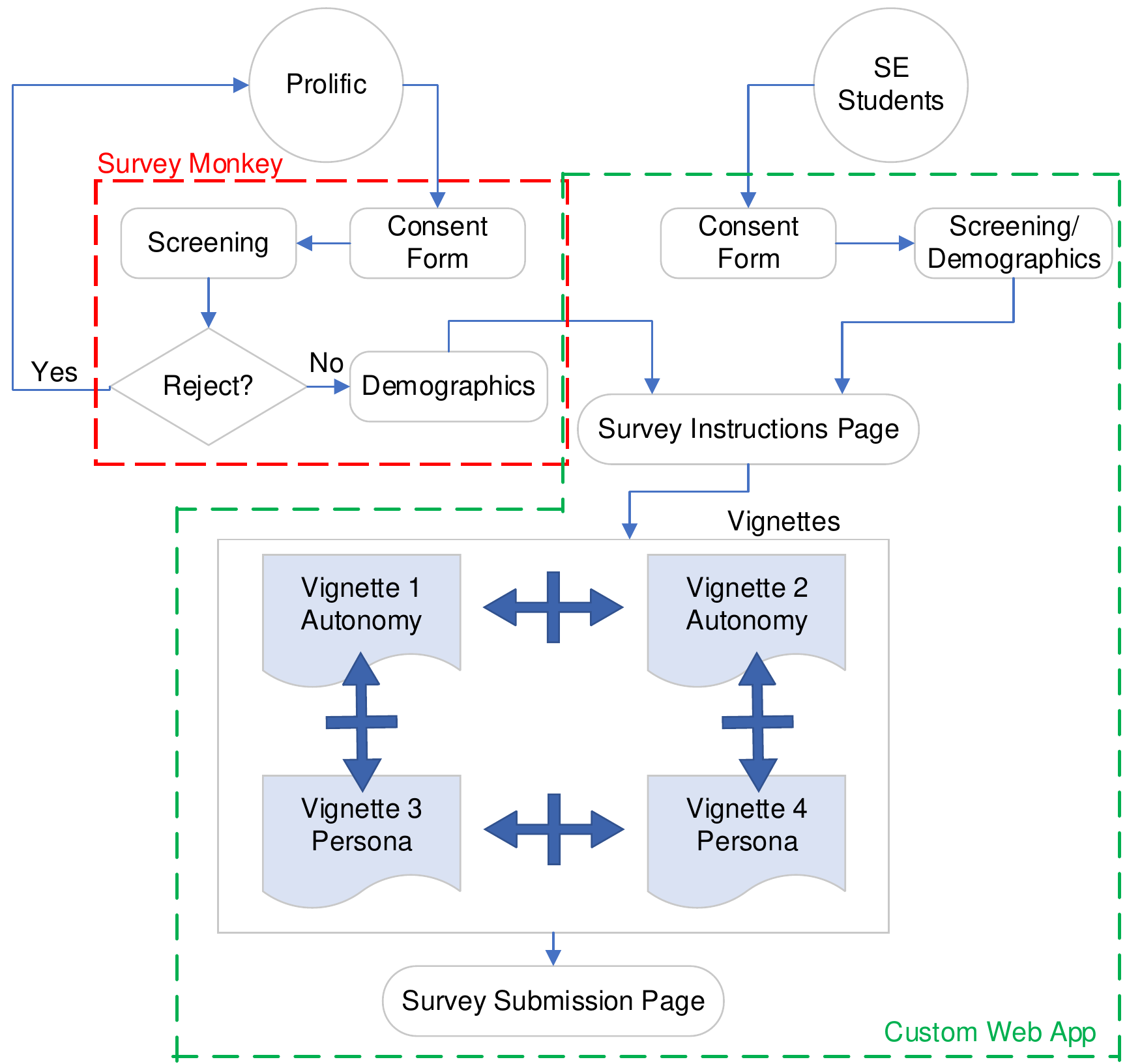}

  \caption {Participant flow in the survey instrument.}
  \label {fig:survey-instrument}

\end{figure}

\noindent
We designed a randomized, vignette-based survey to explore how  \emph{autonomy} and \emph{persona}  affect how bots are perceived by a broader set of developers, \ie how they influence the dependent variable \emph{preference} for how a bot behaves. Vignettes are fictional scenarios in which a bot interacts with users on GitHub and the interaction reflects (alternately) autonomy and persona of the bot.
This design addresses RQ\ref{rq:survey} and is summarized in \cref{fig:survey-instrument}.
A survey offers more control over the questions asked and a broader pool of respondents.
We opt for vignettes experiments embedded in a survey, also known as a factorial survey, rather than a traditional survey as it is supports higher degree of realism, allows for systematically varied descriptions, and feels less monotonous~\cite{azu_jmmss20321}.
Vignette-based surveys are an established sociological instrument introduced by Rossi et al. in the 1970s~\cite{ROSSI1974169} and recently used in software engineering research~\cite{LiawEtAl,mcnamara2018acm,PalombaTFOZS21,SarmentoMSCTP22}.
We discuss the design, deployment and results of the survey below.
\looseness -1

\subsection{Target Population}

\noindent
The target population is software developers with knowledge of pull request workflows. There is no obvious list comprising a sampling frame of such a population~\cite{baltes2020sampling}. We follow a non-probabilistic purposive approach. While this threatens generalizability, in our exploratory context we can use the respondent characteristics in our inferences to explore differences. We targeted Software Engineering third year undergraduates from \uvic{}, a midsized, North American university, and Prolific,\footnote{https://www.prolific.co/} an online participant recruitment platform that provides researchers with access to participants around the world with an internal screening mechanism.

\subsection{Survey Design}

\noindent
The survey began with screening and demographic questions, to ensure that the Prolific respondents are in the target population, as recommended by Danilova \etal and Ebert \etal~\cite{danilova2021really, ebert2022ropes}. The most salient questions are shown in \cref{tbl:survey-questions} and the screening questions are in the replication package (Sect. \ref{sec:replication}).

\begin{figure*}[t]
  \centering
  \includegraphics[
    width = \textwidth,
    clip,
    trim = 4mm 2mm 8mm 5mm
  ] {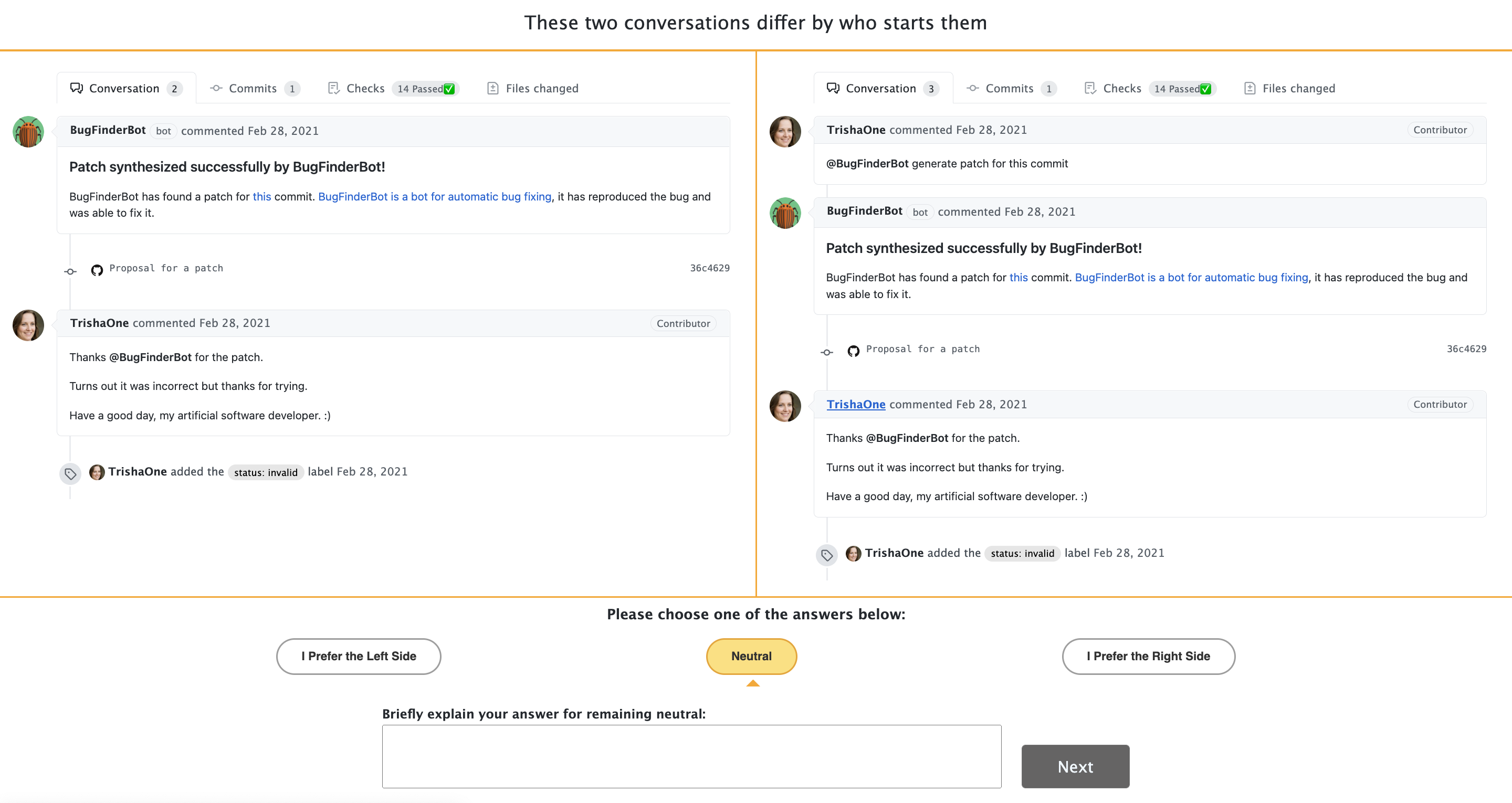} %
  \caption{Screenshot of web application showcasing vignette \#1. User avatar has been redacted for privacy. Bottom shows the choice input and text entry for rationale. Neutral is the default choice.}
  \label{fig:vignette-example}
\end{figure*}

We additionally screened for knowledge of programming and knowledge of pull request based development (e.g., GitHub account, pull request definitions).
Incorrect answers on any screening question meant we ended the survey for that respondent.
For Prolific, we returned the disqualified respondents back to the Prolific Study page to complete the survey, as required by Prolific for compensation. %
These questions were followed by a random ordering of four short \emph{vignettes}~\cite{mcnamara2018acm} representing example bot actions in a GitHub pull request interface (shown in \cref{fig:vignette-example}).
While the themes of autonomy and persona frequently occurred in the interview data, for our survey vignettes we needed to commit to a particular operationalization to create constructs to represent those themes.

\subsubsection*{Constructs---Autonomy}
We define two poles of the autonomy construct, \emph{proactive} and its opposite, \emph{reactive}, based on the codes which emerged in our Phase 1 interview study.
\emph{Proactive} bots can independently take action with no human trigger within a repository.
P7 explained: \emph{``... but when it comes to the lightweight tasks, like doing some really small code changes or some document changes ... Instead, we can automate those tasks by bots''} (P7).
On the opposite pole, \emph{Reactive} bots would first have a human request assistance, \eg the bot is manually triggered by maintainers.
The way in which the Repairnator bot \cite{monperrus2019explainable} independently opens pull requests on repositories is an example of a proactive bot.
A recent study of benchmarking bots has identified four reactive and five proactive bots~\cite{MarkusseLS22}.

\subsubsection*{Constructs---Persona}
We focus solely on the persona as expressed by the bot's textual utterances, leaving neutral the visual appearance of the bot (i.e., we aim to have the bot present a neutral avatar and name). We define two poles of the persona construct. %
A \textit{Factual} bot response uses no humanizing details. It merely delivers the factual content of the message. P10 explained: \emph{``... so, the first thing is if something is correct, and the bot approve it, and it should have a friendly language like, Hey, you did this thing right, and it's fine. And if something is wrong, and then you know, if it rejects the pull request, ideally, it will say me why I'm rejecting the pull request''} (P10). P5 expected a bot to be completely factual; they stated: \emph{``a bot doesn't have any feeling or anything. It is just a piece of software that's telling you to do something''} (P5).
On the opposite pole, a bot that exhibits a human-seeming persona we term \emph{Personable}.
This bot uses informal and very enthusiastic text. For example, P6  stated: \emph{``even some bots can have some characters, right, either a small emoji at the end, that kind of makes a big difference ...  It just gives a little bit of a human factor to the bot''} (P6).
For example, in Vignette 3 the Personable bot first thanks the contributor for their pull-request, and then mentions it is `valuable'.

Our final construct for the survey was the respondent's \textbf{Preference} for the bot's behavior in a given vignette.
Vignettes were constructed to capture both poles of each of our two constructs (\ie Factual-Personable and Proactive-Reactive).
We left the interpretation of preference up to the participants and used an open-ended response to characterize how they justified their preference.
We encode it as a ternary-valued variable: prefer left, neutral, prefer right.

\subsubsection*{Survey Vignettes}
Following the short-answer screening and demographic questions, we presented vignettes.
Vignettes are fictional scenarios in which a bot interacts with users on GitHub and the interaction reflects (alternately) autonomy and persona of the bot.
Each vignette represents a single pull request discussion, extracted from existing GitHub examples for verisimilitude.

We change the bot avatar to appear neutral, and change names of the GitHub contributors to protect their privacy.
We also modify the vignette's order of discussion and/or the bot's messages, to emphasize the construct (autonomy or persona) under test.
This results in two separate scenarios per vignette, as shown in \cref{fig:vignette-example}: on the left, for one extreme of the construct (\eg the proactive bot) and one on the right for the other extreme (\eg the reactive case).

Respondents were prompted with an indication as to what is different between the two sides of the vignettes.
We state ``these two conversations differ by who starts them'' (autonomy) and ``these two conversations differ by the text used by [bot]'' (persona).
We try to phrase this neutrally to avoid biasing respondents to one side or the other. %
This prompting saves respondents time hunting for differences that might only be there accidentally, and focuses them on the construct being studied.
We ask the respondent to carefully examine each scenario in the vignette and then indicate which they prefer: Left, Neither, Right.
Respondents were also required to write why they preferred their chosen side of a scenario.

To ensure the two independent variables do not interact in the vignettes, each vignette varies only in one of the constructs.
For example, in the right-most vignette of \cref{fig:vignette-example} the left and right scenarios show a proactive and reactive
bot interaction respectively.
Persona is fixed to factual in both sides.
In this case, that is easily done by ensuring the text used by the bot is identical in each scenario.
\cref{tbl:survey-questions}, bottom, outlines this in detail.
\looseness = -1

\subsection{Deploying the Survey}

\noindent
We sent out the invitations for this survey in different batches spread out several days and
timestamps as suggested by Ebert \etal~\cite{ebert2022ropes}.
Participants were paid an hourly rate of 7.5\pounds\ to complete the survey, even if they were screened out.
\looseness = -1

In total our survey received $N=56$ valid responses (30 \uvic{}, 26 Prolific).
We released the first iteration of the survey in late February 2022.
Of 69 invited students from \uvic{}, 30 opted in to the survey.
300 Prolific users were invited to the survey over 4 different iterations spread out over roughly one month.
From the 300 invited Prolific users we screened out 274 participants that failed to successfully complete the screening questions,
leaving us with a total of 26 valid Prolific responses.
Such a screening success rate (8.7\%) is not uncommon in studies with a highly technical target populations. This ensures valid participants. 
Danilova \etal saw a rate of 25\% for a simpler technical question~\cite{danilova2021really}.

34 (61\%) of the respondents were students.
19 identified as female, 34 as male, 2 non-binary, and 1 did not disclose.
34 respondents were 18-24, 18 were 25-34, and 4 were 35 or older.
27 had 1-3 years of software development experience, 19 had more than 3 years, and 10 had less than a year.
10 had previous bot interactions. Only 6 could not define a software bot.
\looseness = -1

\begin{table}[th!]
\footnotesize
\caption{Survey questions and vignettes}%
\renewcommand\arraystretch{1.1}
\label{tbl:survey-questions}
  \begin{tabular}{p{.55\columnwidth}p{.35\columnwidth}} %
    \toprule
    \textbf{Demographics} & \textbf{Answers} \\
    \midrule
    To which gender identity do you most identify? & One of: \emph{Male, Female, None-binary, Other, Prefer not to say.}\\
    How old are you? & One of: \emph{Under 18, 18-24, 25-34, 35-44, 45-54, 55-64, 65+} \\
    What is your profession? & \emph{Open input.} \\
    How many years of experience do you have in your profession? & One of: \emph{Less than one year, one to three years, three to five years, more than five years.} \\
    Have you ever submitted a pull request to any open-source project? & One of: \emph{Yes, no.} \\
    Have you ever reviewed pull requests on any open-source project? & One of: \emph{Yes, no.} \\
    How often do you collaborate with repositories on Github, Gitlab, Bitbucket, etc? & One of: \emph{At least once a day, At least once a week, At least once a month, At least once every six months, At least once a year, Never.} \\
    Have you ever interacted with a bot in an open-source project? if yes, please enter the name of at least one of these bots. & \emph{Open input.} \\
    What is a Bot? Briefly describe or provide a definition. & \emph{Open input.} \\
    \midrule
    \textbf{Screening} &\\
    \midrule
    Screening questions are included in the attached replication package (Sect. \ref{sec:replication}). &\\
    \midrule
    \textbf{Vignettes}  & \textbf{Variables} \\
    \midrule
    \textbf{Vignette \#1:} In this vignette the BugFinderBot generates a patch for a commit in a PR. In the left scenario it generates the commit autonomously, in the right scenario it is manually triggered by a developer. & Variable: Autonomy, persona is fixed to factual. \\
    \textbf{Vignette \#2:} The situation in this vignette is similar to vignette \#1, with the only difference being the informal (personable) writing of the bot versus the more factual writing in vignette \#1. & Variable: Autonomy, persona is fixed to personable. \\
    \textbf{Vignette \#3:} A CLA bot responds to a new user opening a PR by asking them to sign a CLA. In the left scenario the bot is factual, in the right it is personable.  & Variable: Persona, Autonomy is fixed to reactive. \\
    \textbf{Vignette \#4:} In this vignette a housekeeper bot responds to a user opening a PR, informing them of the house-rules of the repository and letting them know that the PR targets the wrong branch. & Variable: Persona, Autonomy is fixed to proactive. \\

    \bottomrule
    \end{tabular}
\end{table}

\subsection{Descriptive Analysis and Results}
\label{sec:descriptive}
\noindent
To understand how Autonomy and Persona influence preference (\refrq{rq:survey}),
we begin with descriptive statistics of the results.
We then compare models for our independent variable of preference based on different formulations of possible predictors (including persona, autonomy, and GitHub activity).

\begin{figure}[ht]
  \centering
  \includegraphics*[width=.7\linewidth]{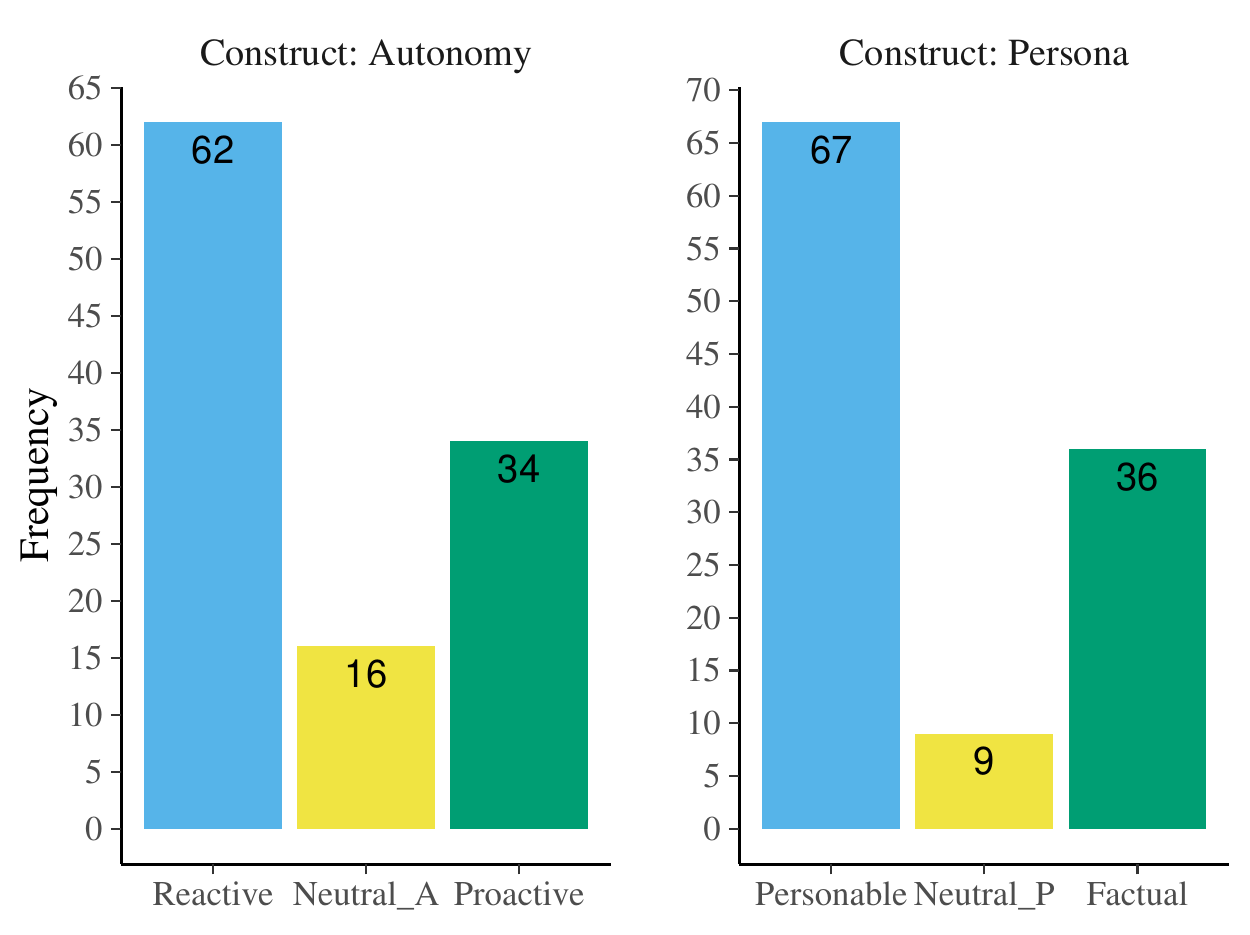}
  \caption{Summary of respondent choices across both vignettes, per construct.}
  \label{fig:descr_overall}
\end{figure}

\cref{fig:descr_overall}
shows the overall choice distributions. Our respondents preferred reactive, personable bots over proactive and factual bots.
If we split out the individual vignettes, as shown in \cref{tab:scenarios_summary}, a slightly different picture arises.
The pattern holds for reactive bots, as in both Vignettes 1 and 2 a strong majority prefer reactive.
However, while in Vignette 3 there is a strong preference for personable (39) over factual (13), in Vignette 4 this advantage weakens (28:23, respectively).

\begin{table}
  \caption[short]{Vignette construct choices. %
  }
\begin{tabular}[t]{clr|clr}
  \toprule
  Vignette & Construct & Count & Vignette & Construct & Count\\
  \midrule
  1 & Reactive & 31 &  3 & Factual & 13 \\
  1 & Neutral & 6 & 3 & Neutral & 4 \\
  1 & Proactive & 19 & 3 & Personable & 39\\\midrule
  2 & Reactive & 31 & 4 & Factual & 23 \\
  2 & Neutral & 10 & 4 & Neutral & 5 \\
  2 & Proactive & 15 &  4 & Personable & 28 \\
  \bottomrule
  \end{tabular}
  \label{tab:scenarios_summary}
\end{table}

We also considered proxies for programming experience.
There are different ways of operationalising programming experience~\cite{DBLP:journals/ese/SiegmundKLAH14}, and we consider several different proxies that might provide complementary insights as to how more experience might impact preferences: GitHub activity levels; pull request experience; and overall experience (in years) programming.
\looseness -1

\begin{figure*}
  \centering
    \includegraphics[width=1\linewidth]{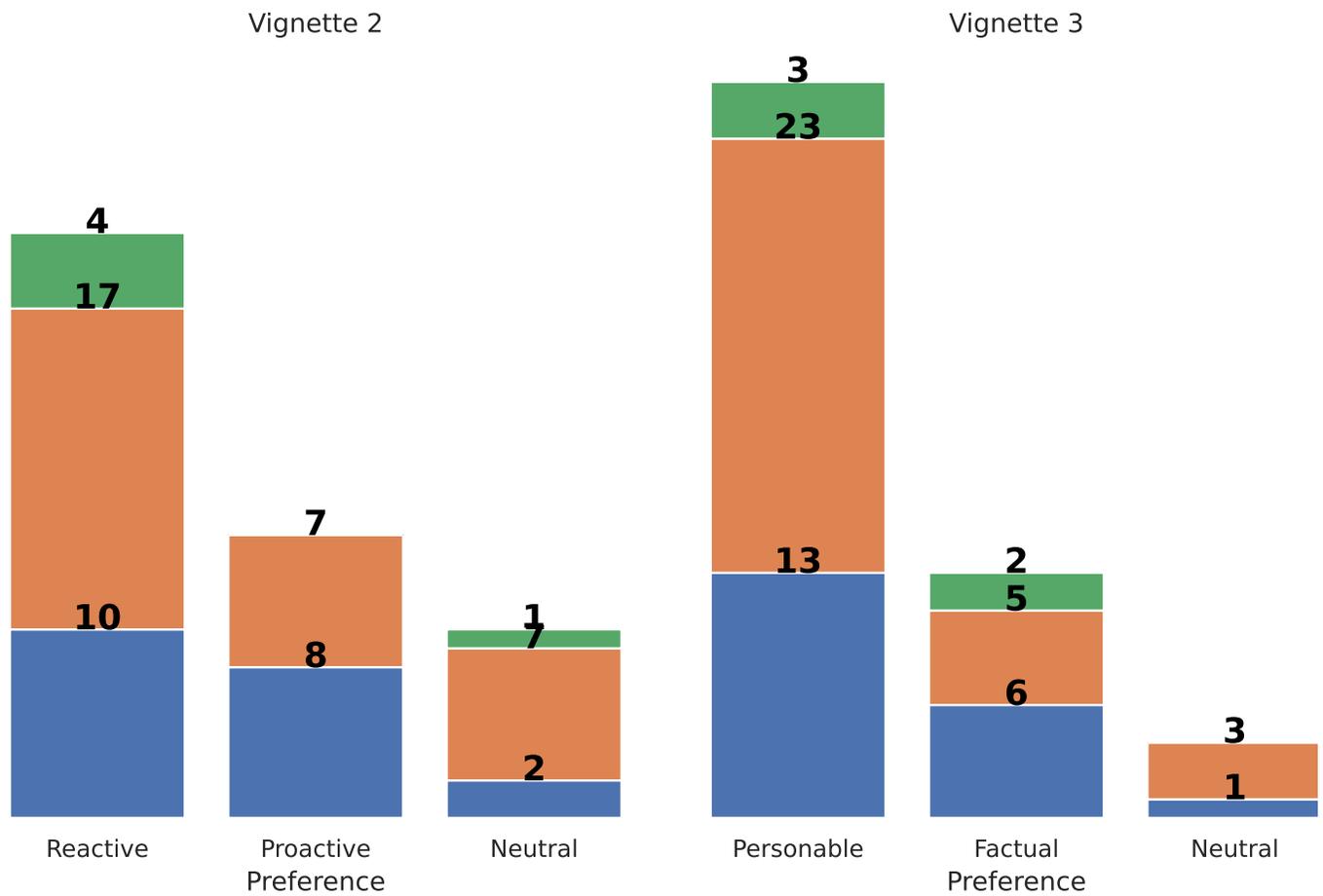}
  \caption{Vignette choice conditioned on GitHub activity.}
  \label{fig:gh_scenarios}
 \end{figure*}

\cref{fig:gh_scenarios} refines the analysis by conditioning on GitHub activity.
Our definition of activity is that
\emph{Experienced} GitHub users use GitHub more frequently than once a month.
\emph{Infrequent} users, although they have a GitHub account, use GitHub less than once a year.
\emph{Intermediate} users fall somewhere in between those extremes.
Interestingly, experts prefer reactive and proactive bots equally (Vignettes 1 and 2), while
novices and intermediate users prefer reactive bots more strongly (Fisher's exact test, p=0.015, $\alpha$=0.05).
Liao \etal~\cite{liao2016personalai} similarly found that expert users are less likely to accept input of automated agents, and are wary of proactive automated agents for fear of interruptions.
\looseness = -1

Also relevant is bot experience. We asked participants to define a bot, and if they had interacted with at least one bot on an open-source project, to list at least one, in \cref{tbl:survey-questions}.
10 respondents gave some bot examples, and 44 did not answer this non-required question.
6 respondents gave a definition of bot which indicated they were unsure.
If participants defined bots clearly, we listed them as having moderate knowledge of bots (40/56). If they also listed previous bots they had interacted with, we classified them as experienced with bots (10). The remainder we categorized as inexperienced (6).
The overall pattern holds across bot experience (prefer Reactive and Personable).
There is no effect of bot experience on Autonomy preference (p=0.27, $\alpha$=0.05) in our data.
There also does not seem to be any particular pattern based on a respondent's pull request activities.
\looseness = -1

\subsection{Explaining Choices With Open-Ended Answers}

\noindent
To better understand how respondents evaluate autonomy and persona we open-code their rationale for their preference (bottom of \cref{fig:vignette-example}).
Three authors of the paper independently coded the open answers
of the respondents in three rounds, and after each round the coders discussed their conflicts to
increase shared understanding of the coding task and improve the coding guide.
We had substantial inter-rater agreement of $\kappa = 0.723$ by the conclusion of the final round.
This resulted in five high-level codes, each with two sub-codes representing opposite polarities (\cref{tbl:codes}).  Code frequencies are shown in parentheses.
We also included a code for comments stating that they have no preference, or for noise/unusable data.

Respondent justifications for preferences were most often because the bot had a Clear message (41 occurrences), the bot was Polite (40), or the message was Information Rich (19). For preferences about bot autonomy, there was an even split between Bot-Initiated (30) and Human-Initiated (41) control.
The open-ended answers showed that our constructs were indeed what participants were reacting to, and echo the themes from the Phase 1 interviews.

\begin{table}[t]
  \caption{Open coding results for the open-ended question on rationale.}
\centering
\begin{tabular}{p{1cm}p{1.3cm}p{5.5cm}}
\toprule
  Code & Polarity (freq.) & Definition \\ \midrule
Noisiness & Noisy (4) / Clear (41) & The bot clutters the conversation by sending too many messages or pinging developers frequently, OR messages are clear, short and easy to understand. \\
  Persona & Polite (40) / Rude (7) & The bot is friendly and the language used by this bot is similar to human conversation OR The bot does not care about human feelings in their messages and actions. \\
  Control & Bot- (30) / Human-initiated (41) & The bot is in control, initiates actions, and is accountable for these actions  OR The developer is in control of the bot, initiates it, and is responsible for the bot's actions \\
  Productivity & Boosts (2) / Hurts (1) & The bot is efficient and improves the workflow by saving time and resources OR The bot is inefficient, consuming resources and wasting time    \\
  Information Content & Rich (19) / Poor (1) &  The bot is configured to provide enough  information for the assigned task OR  The bot does not produce useful information \\ \bottomrule
  \end{tabular}
\label{tbl:codes}
\end{table}

\subsection{Bayesian Analysis of Quantitative Data}

We use Bayesian Data Analysis\,\cite{gelman:2013:bda} to explore possible influences on subject preferences, following the example of Furia \etal~\cite{Furia2022} and the guidelines of Torkar \etal\,\cite{Torkar2021}.
In a frequentist setting we would not be able to find what are possibly small effects with a sample size of 56.
However, Bayesian analysis yields valid predictions even with less data, albeit with the addition of a (in our case, weakly informative) prior probability distribution~\cite{gelman}.

\smallskip

\subsubsection*{Causal Model and Associated Statistical Models}
We created a causal graph~\cite{Rohrer2018} to model the relationships between themes from our interviews (\cref{sec:coding}). A causal model is a Directed Acylic Graph (DAG) in which arrows reflect a causal relationship (an influence) between variables.
Our causal graph is available in our replication package.
The model captures that  \variable{BotExperience}, \variable{GitHubActivity}, and \variable{PRActivity}
influence a preference for degree of Autonomy and Persona style. A hidden node captures other, unmodeled sources of variation.

We codify Preference with the variables \variable{VignetteChoice\_Autonomy} and \variable{VignetteChoice\_Persona}.   The treatment effects ``causing'' these choices are the \variable{AmountOfAutonomy} and \variable{AmountofPersona}.
That is, we model the probability of a respondent's choice (in our instrument ``Prefer Left,'' ``Neutral,'' or ``Prefer Right'') in the Vignettes of \cref{fig:vignette-example}.
\looseness = -1

\begin{table}[b!]

	\caption{Candidate models for Autonomy preferences}%
  \label{tab:models}

  \begin{tabular}{cp{0.7\linewidth}}
  \toprule
  Model Number & Statistical Model \\
  \midrule
  	$\mathcal{M}_1$ & VignetteChoice\_Autonomy $\sim$ GitHubActivity \\
  	$\mathcal{M}_2$ &  VignetteChoice\_Autonomy $\sim$ BotExperience + GitHubActivity + PRActivity + Experience  \\
    $\mathcal{M}_3$ &  VignetteChoice\_Autonomy $\sim$ VignetteChoice\_Persona \\
  	\bottomrule
  \end{tabular}
\end{table}

We derive ordinal regression models~\cite{Brkner_2019} from the causal graph and outline three instances in \cref{tab:models}. In the table, a tilde should be read as ``dependent variable (left side) is explained by independent variable interactions'' (right side). We create 3 models for each of the four vignettes (12 total models). The table reflects phrasing for models for the Autonomy vignettes (1 and 2), but similar expressions are used for Persona (Vignettes 3 and 4).

Model \(\mathcal M_1\) is the simplest. It assumes that the level of \variable{GitHubActivity} alone is a sufficient predictor in each of the vignettes.  Model \(\mathcal M_2\) adds expertise with PRs, bots, and overall software development experience, to check whether these increase the strength of explanation.
Model \(\mathcal M_3\) checks whether the responses are correlated across vignettes for independent constructs (\eg whether one's choice for Autonomy influences choice for Persona, and vice-versa).

The goal of the analysis is to find a model that best explains the data with the fewest predictors.  Adding more predictors to the model raises the risk of over-fitting (biasing) to the collected data and reduces our ability to explain more general phenomena. We then compared the explanatory power of our different models. The more a model can explain the data, the more indicative that is of potentially important effects.

\subsection{Quantitative Analysis Results}

\noindent
We follow the guide of B\"{u}rkner and Vuorre~\cite{Brkner_2019} to perform inference. A cumulative/continuous response ordinal model of a latent parameter $\widetilde{Y}_{\mathrm{autonomy}}$ (resp. $\widetilde{Y}_{\mathrm{persona}}$) models the user's general preference for an aspect of bot behavior.
The latent continuous variable $\widetilde{Y}$ is partitioned into the three responses, ``reactive," ``neutral," ``proactive'' (resp.\ ``factual,'' ``neutral,'' ``personable''). The intuition is that a partitioning properly treats the data as ordered categories rather than a continuous metric response, and the most likely partitioning is found using Bayesian inference.

\smallskip

\subsubsection*{Conditional Effects From the Posterior Distribution}
Posterior inference in the three models finds the distribution of regression parameters (the R scripts for the analysis are found in the replication package, Sect. \ref{sec:replication}). Following the Bayesian workflow of Gelman \etal~\cite{GelmanWorkflow}, elaborated for software research by Torkar \etal \cite{Torkar2021} we use a model comparison approach that focuses on building an adequate explanation for the observed data that can be useful in answering questions, doing decision analysis, or making predictions. It does not imply it is the \emph{best} possible model, just one that given the various factors in our causal model, best explains (`is least surprised by') the data.

The posterior distribution, produced by the inference procedure, allows us to evaluate the probability of a particular choice in our vignettes, conditional on some independent variable.
Our objective is to explore how independent variables might influence our dependent variables, if at all. %

\begin{figure}
  \centering
  \includegraphics[width=.8\linewidth]{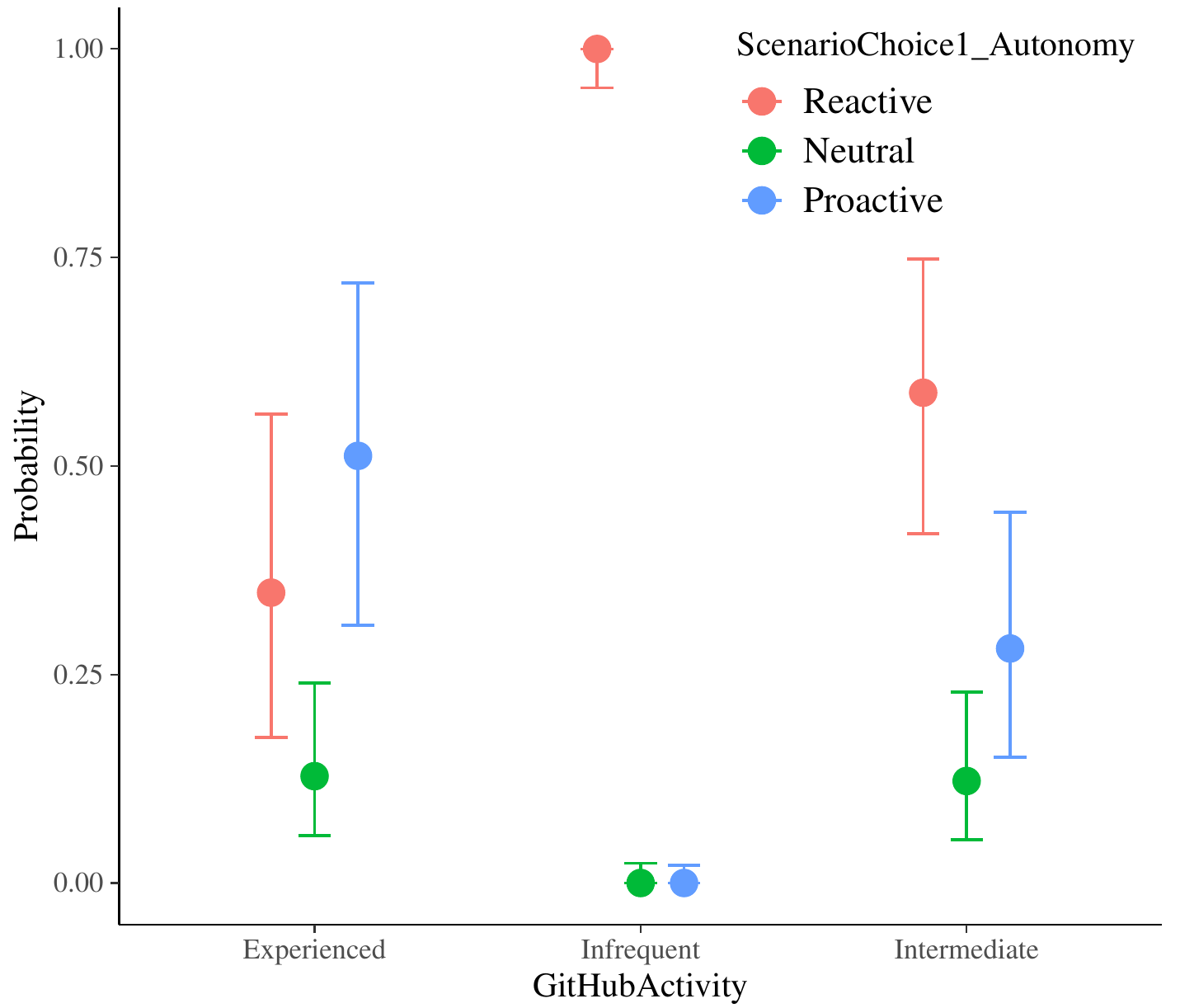}
  \caption{Conditional effects plot showing probability of Vignette 1 choice conditioned on GitHubActivity. From left to right, Reactive (red), Neutral (green), and Proactive (blue).}
  \label{fig:bayes_sc1}
\end{figure}

The inference process generates a posterior predictive distribution, one of the most useful differences with other inference approaches. Here, we can use the posterior to ask questions about the models. %
The conditional effects plot in \cref{fig:bayes_sc1} shows a sample from the posterior. The x-axis captures one of the predictors in the model, in this case the amount of a respondent's GitHub activity (experienced, infrequent, intermediate). On the y-axis we show the probability of the response categories, that is, of choosing either Proactive (blue dots), Neutral (green dots), or Reactive (red dots) in Vignette 1. Error bars capture the 95\% credible intervals from the model (i.e., of the samples drawn from the posterior, 95\% fall within the error bars).  %
The dot represents the mean of the samples taken.
\looseness -1

In this vignette
$\mathcal{M}_1$ shows that intermediate and infrequent GitHub users are more likely to choose reactive (blue/rightmost means), while experienced users are more likely to choose proactive bot actions in the Vignette.
Overlapping error bars indicate lower confidence in the probability of the choices of the experienced users, however.

\subsubsection*{Model Comparison with LOOIC}
We can also use information criteria and leave-one-out validation (LOOIC, \cite{Vehtari2016}) to approximate the likelihood of the held-out data based on the observed data and quantify which model is most informative.

A lower LOO Information Criterion score (LOOIC) indicates that the model is a better fit for the data~\cite{Furia2022}.
The relative differences of the score for different models of the same data can be compared. Where LOOIC scores are less than an integer multiple of the standard error, other factors such as our domain knowledge (\ie how likely is a predictor to influence the result) and model parsimony (fewer predictors are preferred over more) are also important.

\begin{figure}
  \centering
  \small
\begin{tabular}{c|ccc|ccc}
  \toprule
  Model & Vignette & LOOIC & SE & Vignette & LOOIC & SE \\
  \midrule
  $\mathcal{M}_1$ & \multirow{3}{*}{V1} & \textbf{101.4}  & 8.7 & \multirow{3}{*}{V2} & \textbf{117.8} & 9.1  \\
  $\mathcal{M}_2$  & & 116.2 & 12.1 & &  120.5 & 12.2 \\
  $\mathcal{M}_3$  & & 111.9 & 8.7   & & 119.4 & 7.6  \\
  \midrule
  $\mathcal{M}_1$ & \multirow{3}{*}{V3} &  95.5 & 11.2 & \multirow{3}{*}{V4} & 112.3  & 10.8 \\
  $\mathcal{M}_2$ & & 115.7 & 14.1 && 127.6  & 7.3 \\
  $\mathcal{M}_3$ & & \textbf{87.6} & 10.6 & & \textbf{111.1} & 7.7 \\
  \bottomrule
\end{tabular}

  \caption[]{Comparing models using Leave-One-Out Information Criterion. Lower values are better. SE is standard error. Model 1 is preferred for vignettes 1 and 2 (autonomy), while Model 3 is preferred for vignettes 3 and 4 (Persona).}%
  \label{tab:looic}

\end{figure}

Our model reflecting prior experience with software development or bots ($\mathcal{M}_2$) was less informative, with wider margins of error, as shown in \cref{tab:looic}. Choices on the Persona construct (Vignettes 3 and 4) were more likely to be influenced by choices on Vignette 1 (\ie whether respondents preferred the reactive or proactive bot actions).

Indeed, returning to \refrq{rq:survey}, the inferred model confirms that inexperienced developers prefer reactive and personable bots. This is consistent with reports elsewhere that that inexperienced developers perceive bots as automated, smart procedures (Erlenhov's Sam persona~\cite{erlenhov2019current}). The models lend evidence to our insight that experienced developers prefer more proactive, factual bots.
\looseness -1

\section{Discussion}

\subsection{Autonomous Bots and Developer Perceptions}

\noindent
Users and professionals seem to be wary of proactive agents~\cite{liao2016personalai,seiffer2021understanding,schaffer2019icandobetter}.
Wessel \etal highlighted that overzealous bots in open-source are perceived by developers as noise~\cite{wessel2021dont}.
Our study found that more experienced developers are more inclined to prefer proactive bots compared
to the less experienced developers. %
The reasons given by our participants only rarely include Wessel \etal's notion of bot noise or
interruptions.
Instead, the developers confidence in (or tolerance for) proactive bots seems to grow as developers become more
experienced. Conversely, less experienced developers may find proactive bots either too intimidating or assertive.
Our study shows that not only should bot creators decide how their bots should be configured, but also reminds us that the developer is and remains central to the development process. Tools which assume otherwise may see little adoption.
Therefore we propose that tuning the levels of bot autonomy depending on the developer's level of experience should be considered in bot design.
Therefore, we propose:

\begin{leftbar}
     \noindent \textbf{Recommendation \#1}: Developers should be given options to scale the autonomy of the bots involved in their development process. For instance, they could choose between \emph{Proactive} and \emph{Reactive}.
\end{leftbar}

\noindent
Developers might also choose between bots with which they feel comfortable (such as Dependabot, which \eg P10 found ``very useful'') and those they prefer to mute.
However, this tuning should be based on developer preferences, as captured by our survey and that of Erlenhov \etal~\cite{erlenhov2020empirical}. Therefore,
\begin{leftbar}
    \noindent\textbf{Recommendation \#2}: In order to tune bot behaviors, projects must have a better way of getting feedback on bot behavior and developer preferences.
\end{leftbar}

\noindent
One way to do this is to capture developer reactions to bot comments, \eg using a thumbs-up emoji.

\subsection{The Importance of Bot Persona}
\noindent
The language used by bots shapes how bots are perceived by their users~\cite{clark2019goodconvo}.
Volkel \etal found that users of bots might be more likely to accept bots that reflect their own personality~\cite{volkel2019understanding}.
Our findings show that developer preference for persona depends on the specific context of a vignette.
The practical implication is that persona matters and should not be an afterthought.
Developers should be given options to influence the persona of the bots involved in their development activities, especially since it is known that bots which display higher levels of anthrophormism build more trust with developers~\cite{seiffer2021understanding, li2021machinelike,pfeuffer2019anthropomorphic}.
Developers could for instance select a persona from an array of predefined personas and styles of communication. For example, some interviewees would rather have a bot stick to the information in the message and skip non-informative utterances. Based on these insights, we propose:

\begin{leftbar}
     \noindent  \textbf{Recommendation \#3:} Developers should be able to select and change the persona of a bot based on their\linebreak[4] own preferences.
\end{leftbar}

\noindent
Future work should look at how other factors influence developers preference for bots' personas.
For example, open source project particularities could be used to guide the design of pre-defined personas inline with  community preferences.
Alami \etal \cite{alami2020foss, alami2021pull} found that open source communities PR governance process tend to align to three distinct but not exclusive styles of governance: lenient, equitable, and protective.
Different bot persona styles that align with community styles could be investigated as potential default choices for contributors.

\subsection{Broader Themes in Bot Interactions}
\noindent
Our study focused on the key themes of autonomy and persona, and implies that there should be more bot configurability and awareness of developer perceptions.
There are other dimensions to developer preferences on bots, however.
Emerging in our interviews and survey responses were themes around
trust, past bot experiences, appearance, and identity. To further understand the roles bots play in
software development these themes need more extensive study.
As complex interactions between software bots and developers expand, a broader set of theories and results around behavior, trust, and language will be needed.
The notion of social norms from Bicchieri~\cite{norms}, Searle's speech acts~\cite{Searle1969SpeechAA}, or politeness theory from Brown and Levinson~\cite{Brown1987PolitenessSU}, are some examples which make this area a fruitful one for transdisciplinary research.

\section{Trustworthiness and Validity of Findings}
\label{sec:threats}

\noindent
\emph{Trustworthiness:}
Trustworthiness assesses the validity of our qualitative analyses and conclusions.

Credibility \& confirmability. We used robust labeling with multiple authors and clear guidelines to reduce bias of a single labeler. We reported earlier on rater reliability for our coding of the survey responses. 

Our sampling technique in Phase I might be seen as a potential weakness of our study. Participants for the interviews were found via referrals and word of mouth. As a consequence, it is feasible that our sample may skew the Phase I findings. However, our mixed-methods approach addressed this weakness somewhat; in Phase II, we surveyed a wide sample of people from a range of experiential, and national backgrounds.

Dependability. Our codebook and the anonymized transcripts and answers are available in our replication package for reliability checks of our findings.

Transferability. We consider we have met the transferability standards by demonstrating that the study results might be used in other comparable circumstances (i.e., free and open source communities). A replication package that is thorough enough to enable other researchers to duplicate a comparable investigation is provided. We also assured transferability by properly defining the study, explaining the participants' various experiences, executing methodology, and evaluating the outcomes in a quantitative phase with a broader sample.

\emph{Internal Validity:}
We piloted the survey with three members of our lab. 
Designing the survey involved trade-offs. 
We randomized the order of the vignettes per participant, and we kept some questions simpler than might be desirable, in order to reduce experimental bias such as fatigue.
We changed the context of each vignette to ensure there was no learning effect.
We keep the amount of text approximately the same on both sides of a vignette.
We checked correlation between vignette choices to ensure vignette choices were correlated on the same construct, but not correlated between constructs. 

In both Vignette 1 \& 2 the bot's PR is rejected. This failure might have influenced how developers 
prefer or perceive acceptability of the autonomous bot. 
However, we did not see any mention of the bot failure in the justifications provided by respondents for preferring the reactive bot.
In our Bayesian modeling, we might have too few novice samples to draw on, resulting in lower model efficiency. 
This is apparent from the larger standard error for these categories.

\emph{Construct Validity:}
Our findings were based on operationalizations of the constructs of persona and autonomy, mapped onto the developer's preference for a bot's behavior. 
These are complex and often personal constructs.
Our operationalization of these constructs necessarily excluded some of this complexity (a necessary trade-off for more control).
Our open-ended questions were an attempt to allow respondents to express this complexity.

To measure preference we asked respondents to indicate which scenario within a vignette a respondent preferred. 
By showing scenarios side-by-side respondents could easily evaluate which pole of the construct shown in the two
scenarios they prefer. 
Additionally, we asked for open-ended justifications to ensure we understood the rationale behind their preference and 
from the open-ended answers we conclude that respondents state similar reasons as the constructs under test. 
We had mixed results with Vignette 4 which might reflect poor phrasing of the interaction. Tuning a bot's persona between factual and personable is a balancing act.
For the modeling we assume that preferring one side is equivalent to not preferring the other side, \ie there are no category-specific effects. 

\emph{External Validity:}
Survey respondents were carefully screened for basic software development knowledge and pull request use.
In addition, our survey samples were drawn from self-described software industry professionals (Prolific), polled at varying times, as well as third year software engineering students. 
The usage of screening 
questions and using several Prolific iterations is in line with current recommendations~\cite{ebert2022ropes,danilova2021really}.
There may be a bias in the people willing to enroll in Prolific studies, and we did not have many participants who were older than 35 or with more than 5 years of experience.

\section{Conclusion}
\label{sec:conclusion}
\noindent
This paper examined the factors influencing developer perceptions of GitHub bots.
We began with an interview study to elicit some important themes in how projects and developers use and perceive software bots.
From this two important constructs emerged, which influenced perceptions: the degree of Autonomy a bot exhibited, and its Persona.
A vignette-based survey allowed us to control other aspects and explore how respondents reacted to bots in a simulated PR discussion.
Participants
with less software experience generally preferred reactive, less autonomous bots, but tolerated friendly personas.
Bot development for software projects should make these two aspects of bots more configurable. Better understanding of project and developer interactions is important for making full use of bots.

\paragraph*{Replication Package}\label{sec:replication}

\noindent We made the study's %
{data and other artifacts}\footnote{\url{https://doi.org/10.5281/zenodo.7040317}} openly available. For Phase I, we shared the full and detailed interview guide, anonymized interview transcripts, and the result of the data analysis, i.e., codebook. For Phase II, the survey design and data.

\paragraph*{Human Research Ethics Review}\label{sec:hreb}
The study, including the power-over relationship with students, was approved by the relevant institutional review boards of the researchers involved.

\paragraph*{Acknowledgments}
\noindent We thank the interview and the survey participants for their time and useful responses. We also thank Rohith Pudari for his help with vignette tooling.

\balance
\bibliographystyle{IEEETranS}
\bibliography{bots}

\begin{thebibliography}{10}
\providecommand{\url}[1]{#1}
\csname url@samestyle\endcsname
\providecommand{\newblock}{\relax}
\providecommand{\bibinfo}[2]{#2}
\providecommand{\BIBentrySTDinterwordspacing}{\spaceskip=0pt\relax}
\providecommand{\BIBentryALTinterwordstretchfactor}{4}
\providecommand{\BIBentryALTinterwordspacing}{\spaceskip=\fontdimen2\font plus
\BIBentryALTinterwordstretchfactor\fontdimen3\font minus
  \fontdimen4\font\relax}
\providecommand{\BIBforeignlanguage}[2]{{%
\expandafter\ifx\csname l@#1\endcsname\relax
\typeout{** WARNING: IEEEtranS.bst: No hyphenation pattern has been}%
\typeout{** loaded for the language `#1'. Using the pattern for}%
\typeout{** the default language instead.}%
\else
\language=\csname l@#1\endcsname
\fi
#2}}
\providecommand{\BIBdecl}{\relax}
\BIBdecl

\bibitem{empiricalstandards}
``Supplements | empirical standards,''
  \url{https://acmsigsoft.github.io/EmpiricalStandards/Supplements/?supplement=InterRaterReliabilityAndAgreement},
  (Accessed on 01/03/2023).

\bibitem{alami2020foss}
\BIBentryALTinterwordspacing
A.~Alami, M.~L. Cohn, and A.~W{\k{a}}sowski, ``How do {FOSS} communities decide
  to accept pull requests?'' in \emph{Proceedings of the Evaluation and
  Assessment in Software Engineering}.\hskip 1em plus 0.5em minus 0.4em\relax
  {ACM}, apr 2020. [Online]. Available:
  \url{https://doi.org/10.1145%2F3383219.3383242}
\BIBentrySTDinterwordspacing

\bibitem{alami2021pull}
A.~Alami, R.~Pardo, M.~L. Cohn, and A.~Wasowski, ``Pull request governance in
  open source communities,'' \emph{IEEE Transactions on Software Engineering},
  2021.

\bibitem{baltes2020sampling}
\BIBentryALTinterwordspacing
S.~Baltes and P.~Ralph, ``Sampling in software engineering research: a critical
  review and guidelines,'' \emph{Empirical Software Engineering}, vol.~27,
  p.~94, 7 2022. [Online]. Available:
  \url{https://link.springer.com/10.1007/s10664-021-10072-8}
\BIBentrySTDinterwordspacing

\bibitem{norms}
C.~Bicchieri, \emph{Norms in the Wild: how to Diagnose, Measure and Change
  Social Norms}.\hskip 1em plus 0.5em minus 0.4em\relax Oxford University
  Press, 2016.

\bibitem{Brown1987PolitenessSU}
P.~E. Brown and S.~Levinson, \emph{Politeness: some universals in language
  usage}.\hskip 1em plus 0.5em minus 0.4em\relax Cambridge University Press,
  1987.

\bibitem{Brkner_2019}
\BIBentryALTinterwordspacing
P.-C. B{\"u}rkner and M.~Vuorre, ``Ordinal regression models in psychology: A
  tutorial,'' \emph{Advances in Methods and Practices in Psychological
  Science}, vol.~2, no.~1, pp. 77--101, feb 2019. [Online]. Available:
  \url{https://doi.org/10.1177%2F2515245918823199}
\BIBentrySTDinterwordspacing

\bibitem{chaves2022chatbotdesign}
A.~P. Chaves, J.~Egbert, T.~Hocking, E.~Doerry, and M.~A. Gerosa, ``Chatbots
  language design: The influence of language variation on user experience with
  tourist assistant chatbots,'' \emph{ACM Transactions on Computer-Human
  Interaction}, vol.~29, pp. 1--38, 4 2022.

\bibitem{chaves2020chatbot}
\BIBentryALTinterwordspacing
A.~P. Chaves and M.~A. Gerosa, ``How should my chatbot interact? a survey on
  social characteristics in human{\textendash}chatbot interaction design,''
  \emph{International Journal of Human{\textendash}Computer Interaction}, pp.
  1--30, Nov. 2020. [Online]. Available:
  \url{https://doi.org/10.1080/10447318.2020.1841438}
\BIBentrySTDinterwordspacing

\bibitem{clark2019goodconvo}
L.~Clark, N.~Pantidi, O.~Cooney, P.~Doyle, D.~Garaialde, J.~Edwards,
  B.~Spillane, E.~Gilmartin, C.~Murad, C.~Munteanu, V.~Wade, and B.~R. Cowan,
  ``What makes a good conversation?'' in \emph{Proceedings of the 2019 CHI
  Conference on Human Factors in Computing Systems}.\hskip 1em plus 0.5em minus
  0.4em\relax ACM, 2019, pp. 1--12.

\bibitem{creswell2017designing}
J.~W. Creswell and V.~L.~P. Clark, \emph{Designing and Conducting Mixed Methods
  Research}, 3rd~ed.\hskip 1em plus 0.5em minus 0.4em\relax SAGE Publications,
  2017.

\bibitem{danilova2021really}
A.~Danilova, A.~Naiakshina, S.~Horstmann, and M.~Smith, ``Do you really code?
  designing and evaluating screening questions for online surveys with
  programmers,'' in \emph{2021 IEEE/ACM 43rd International Conference on
  Software Engineering (ICSE)}.\hskip 1em plus 0.5em minus 0.4em\relax IEEE, 5
  2021, pp. 537--548.

\bibitem{ebert2022ropes}
F.~Ebert, A.~Serebrenik, C.~Treude, N.~Novielli, and F.~Castor, ``On recruiting
  experienced github contributors for interviews and surveys on prolific,'' in
  \emph{1st International Workshop on Recruiting Participants for Empirical
  Software Engineering (RoPES'22)}, 2022.

\bibitem{erlenhov2019current}
\BIBentryALTinterwordspacing
L.~Erlenhov, F.~G. de~Oliveira~Neto, R.~Scandariato, and P.~Leitner, ``Current
  and future bots in software development,'' in \emph{Proceedings of the 1st
  International Workshop on Bots in Software Engineering}.\hskip 1em plus 0.5em
  minus 0.4em\relax IEEE Press, 2019, pp. 7–--11. [Online]. Available:
  \url{https://doi.org/10.1109/BotSE.2019.00009}
\BIBentrySTDinterwordspacing

\bibitem{erlenhov2020empirical}
L.~Erlenhov, F.~G. d.~O. Neto, and P.~Leitner, ``An empirical study of bots in
  software development: Characteristics and challenges from a practitioner’s
  perspective,'' in \emph{Proceedings of the 28th ACM Joint Meeting on European
  Software Engineering Conference and Symposium on the Foundations of Software
  Engineering}, 2020, pp. 445--455.

\bibitem{farah2021conveying}
J.~C. Farah, V.~Sharma, S.~Ingram, and D.~Gillet, ``Conveying the perception of
  humor arising from ambiguous grammatical constructs in human-chatbot
  interaction,'' in \emph{Proceedings of the 9th International Conference on
  Human-Agent Interaction}, 2021, pp. 257--262.

\bibitem{Furia2022}
\BIBentryALTinterwordspacing
C.~A. Furia, R.~Torkar, and R.~Feldt, ``Applying bayesian analysis guidelines
  to empirical software engineering data: The case of programming languages and
  code quality,'' \emph{{ACM} Transactions on Software Engineering and
  Methodology}, vol.~31, no.~3, pp. 1--38, Jul. 2022. [Online]. Available:
  \url{https://doi.org/10.1145/3490953}
\BIBentrySTDinterwordspacing

\bibitem{furnham1986response}
A.~Furnham, ``Response bias, social desirability and dissimulation,''
  \emph{Personality and individual differences}, vol.~7, no.~3, pp. 385--400,
  1986.

\bibitem{gelman:2013:bda}
A.~Gelman, J.~B. Carlin, H.~S. Stern, D.~B. Dunson, A.~Vehtari, and D.~B.
  Rubin, \emph{Bayesian data analysis}.\hskip 1em plus 0.5em minus 0.4em\relax
  CRC press, 2013.

\bibitem{gelman}
A.~Gelman, J.~Hill, and A.~Vehtari, \emph{Regression and Other Stories}.\hskip
  1em plus 0.5em minus 0.4em\relax Cambridge University Press, 2021.

\bibitem{GelmanWorkflow}
\BIBentryALTinterwordspacing
A.~Gelman, A.~Vehtari, D.~Simpson, C.~C. Margossian, B.~Carpenter, Y.~Yao,
  L.~Kennedy, J.~Gabry, P.-C. Bürkner, and M.~Modrák, ``Bayesian workflow,''
  2020. [Online]. Available: \url{https://arxiv.org/abs/2011.01808}
\BIBentrySTDinterwordspacing

\bibitem{jain2018evaluatingchatbots}
M.~Jain, P.~Kumar, R.~Kota, and S.~N. Patel, ``Evaluating and informing the
  design of chatbots,'' in \emph{Proceedings of the 2018 Designing Interactive
  Systems Conference}.\hskip 1em plus 0.5em minus 0.4em\relax ACM, 6 2018, pp.
  895--906.

\bibitem{li2021machinelike}
M.~Li and A.~Suh, ``Machinelike or humanlike? a literature review of
  anthropomorphism in ai-enabled technology,'' in \emph{Proceedings of the 54th
  Hawaii International Conference on System Sciences}, 2021, p. 4053.

\bibitem{liao2016personalai}
Q.~V. Liao, M.~Davis, W.~Geyer, M.~Muller, and N.~S. Shami, ``What can you
  do?'' in \emph{Proceedings of the 2016 ACM Conference on Designing
  Interactive Systems}.\hskip 1em plus 0.5em minus 0.4em\relax ACM, 6 2016, pp.
  264--275.

\bibitem{LiawEtAl}
\BIBentryALTinterwordspacing
S.-T. Liaw, E.~Deveny, I.~Morrison, and B.~Lewis, ``Clinical, information and
  business process modeling to promote development of safe and flexible
  software,'' \emph{Health Informatics Journal}, vol.~12, no.~3, pp. 199--211,
  2006, pMID: 17023408. [Online]. Available:
  \url{https://doi.org/10.1177/1460458206066772}
\BIBentrySTDinterwordspacing

\bibitem{MarkusseLS22}
\BIBentryALTinterwordspacing
F.~Markusse, P.~Leitner, and A.~Serebrenik, ``Using benchmarking bots for
  continuous performance assessment,'' \emph{{IEEE} Softw.}, vol.~39, no.~5,
  pp. 50--55, 2022. [Online]. Available:
  \url{https://doi.org/10.1109/MS.2022.3184430}
\BIBentrySTDinterwordspacing

\bibitem{mcnamara2018acm}
A.~McNamara, J.~Smith, and E.~Murphy-Hill, ``Does {ACM}’s code of ethics
  change ethical decision making in software development?'' in
  \emph{Proceedings of the 2018 26th ACM Joint Meeting on European Software
  Engineering Conference and Symposium on the Foundations of Software
  Engineering}.\hskip 1em plus 0.5em minus 0.4em\relax ACM, 10 2018, pp.
  729--733.

\bibitem{miles1984qualitative}
M.~B. Miles, M.~Huberman, and J.~Saldana, \emph{Qualitative data analysis: A
  methods sourcebook}.\hskip 1em plus 0.5em minus 0.4em\relax SAGE
  Publications, Incorporated, 2013.

\bibitem{mirhosseini2017automated}
S.~Mirhosseini and C.~Parnin, ``Can automated pull requests encourage software
  developers to upgrade out-of-date dependencies?'' \emph{Proceedings of the
  32nd IEEE/ACM International Conference on Automated Software Engineering},
  pp. 84--94, 2017.

\bibitem{monperrus2019explainable}
M.~Monperrus, ``Explainable software bot contributions: Case study of automated
  bug fixes,'' in \emph{Proceedings of 2019 IEEE/ACM International Workshop on
  Bots in Software Engineering (BotSE)}, 2019.

\bibitem{nass1994casa}
C.~Nass, J.~Steuer, and E.~R. Tauber, ``Computers are social actors,'' in
  \emph{Conference companion on Human factors in computing systems - CHI
  '94}.\hskip 1em plus 0.5em minus 0.4em\relax ACM Press, 1994, p. 204.

\bibitem{PalombaTFOZS21}
\BIBentryALTinterwordspacing
F.~Palomba, D.~A. Tamburri, F.~A. Fontana, R.~Oliveto, A.~Zaidman, and
  A.~Serebrenik, ``Beyond technical aspects: How do community smells influence
  the intensity of code smells?'' \emph{{IEEE} Trans. Software Eng.}, vol.~47,
  no.~1, pp. 108--129, 2021. [Online]. Available:
  \url{https://doi.org/10.1109/TSE.2018.2883603}
\BIBentrySTDinterwordspacing

\bibitem{patton2014qualitative}
M.~Q. Patton, \emph{Qualitative research \& evaluation methods: Integrating
  theory and practice}.\hskip 1em plus 0.5em minus 0.4em\relax Sage
  publications, 2014.

\bibitem{peng2019exploring}
\BIBentryALTinterwordspacing
Z.~Peng and X.~Ma, ``Exploring how software developers work with mention bot in
  {GitHub},'' \emph{{CCF} Transactions on Pervasive Computing and Interaction},
  vol.~1, no.~3, pp. 190--203, sep 2019. [Online]. Available:
  \url{https://doi.org/10.1007%2Fs42486-019-00013-2}
\BIBentrySTDinterwordspacing

\bibitem{pfeuffer2019anthropomorphic}
N.~Pfeuffer, A.~Benlian, H.~Gimpel, and O.~Hinz, ``Anthropomorphic information
  systems,'' \emph{Business \& Information Systems Engineering}, vol.~61,
  no.~4, pp. 523--533, 2019.

\bibitem{Rohrer2018}
J.~M. Rohrer, ``Thinking clearly about correlations and causation: Graphical
  causal models for observational data,'' \emph{Advances in Methods and
  Practices in Psychological Science}, vol.~1, no.~1, pp. 27--42, Jan. 2018.

\bibitem{ROSSI1974169}
\BIBentryALTinterwordspacing
P.~H. Rossi, W.~A. Sampson, C.~E. Bose, G.~Jasso, and J.~Passel, ``Measuring
  household social standing,'' \emph{Social Science Research}, vol.~3, no.~3,
  pp. 169--190, 1974. [Online]. Available:
  \url{https://www.sciencedirect.com/science/article/pii/0049089X74900118}
\BIBentrySTDinterwordspacing

\bibitem{Santhanam2022}
\BIBentryALTinterwordspacing
S.~Santhanam, T.~Hecking, A.~Schreiber, and S.~Wagner, ``Bots in software
  engineering: a systematic mapping study,'' \emph{{PeerJ} Computer Science},
  vol.~8, p. e866, Feb. 2022. [Online]. Available:
  \url{https://doi.org/10.7717/peerj-cs.866}
\BIBentrySTDinterwordspacing

\bibitem{SarmentoMSCTP22}
\BIBentryALTinterwordspacing
C.~Sarmento, T.~Massoni, A.~Serebrenik, G.~Catolino, D.~A. Tamburri, and
  F.~Palomba, ``Gender diversity and community smells: {A} double-replication
  study on {B}razilian software teams,'' in \emph{{IEEE} International
  Conference on Software Analysis, Evolution and Reengineering, {SANER} 2022,
  Honolulu, HI, USA, March 15-18, 2022}.\hskip 1em plus 0.5em minus 0.4em\relax
  {IEEE}, 2022, pp. 273--283. [Online]. Available:
  \url{https://doi.org/10.1109/SANER53432.2022.00043}
\BIBentrySTDinterwordspacing

\bibitem{schaffer2019icandobetter}
J.~Schaffer, J.~O'Donovan, J.~Michaelis, A.~Raglin, and T.~Höllerer, ``I can
  do better than your {AI},'' in \emph{Proceedings of the 24th International
  Conference on Intelligent User Interfaces}.\hskip 1em plus 0.5em minus
  0.4em\relax ACM, 3 2019, pp. 240--251.

\bibitem{Searle1969SpeechAA}
J.~R. Searle, ``Speech acts: An essay in the philosophy of language,''
  \emph{Language}, vol.~46, p. 217, 1969.

\bibitem{seiffer2021understanding}
A.~Seiffer, U.~Gnewuch, and A.~Maedche, ``Understanding employee responses to
  software robots: A systematic literature,'' in \emph{International Conference
  on Information Systems (ICIS)}, vol. 2021, 2021.

\bibitem{DBLP:journals/ese/SiegmundKLAH14}
\BIBentryALTinterwordspacing
J.~Siegmund, C.~K{\"{a}}stner, J.~Liebig, S.~Apel, and S.~Hanenberg,
  ``Measuring and modeling programming experience,'' \emph{Empir. Softw. Eng.},
  vol.~19, no.~5, pp. 1299--1334, 2014. [Online]. Available:
  \url{https://doi.org/10.1007/s10664-013-9286-4}
\BIBentrySTDinterwordspacing

\bibitem{azu_jmmss20321}
\BIBentryALTinterwordspacing
P.~Steiner, C.~Atzmüller, and D.~Su, ``Designing valid and reliable vignette
  experiments for survey research: A case study on the fair gender income
  gap,'' \emph{Journal of Methods and Measurement in the Social Sciences},
  vol.~7, no.~2, pp. 52--94, 2017. [Online]. Available:
  \url{https://journals.uair.arizona.edu/index.php/jmmss/article/view/20321}
\BIBentrySTDinterwordspacing

\bibitem{storey2016disrupting}
M.-A. Storey and A.~Zagalsky, ``Disrupting developer productivity one bot at a
  time,'' in \emph{Proceedings of the 2016 24th ACM SIGSOFT International
  Symposium on Foundations of Software Engineering}, 2016, pp. 928--931.

\bibitem{Torkar2021}
\BIBentryALTinterwordspacing
R.~Torkar, C.~A. Furia, R.~Feldt, F.~G. de~Oliveira~Neto, L.~Gren, P.~Lenberg,
  and N.~A. Ernst, ``A method to assess and argue for practical significance in
  software engineering,'' \emph{{IEEE} Transactions on Software Engineering},
  pp. 1--1, 2021. [Online]. Available:
  \url{https://doi.org/10.1109/tse.2020.3048991}
\BIBentrySTDinterwordspacing

\bibitem{vanover2021analyzing}
C.~Vanover, P.~Mihas, and J.~Salda{\~n}a, \emph{Analyzing and interpreting
  qualitative research: After the interview}.\hskip 1em plus 0.5em minus
  0.4em\relax SAGE Publications, 2021.

\bibitem{Vehtari2016}
\BIBentryALTinterwordspacing
A.~Vehtari, A.~Gelman, and J.~Gabry, ``Practical {B}ayesian model evaluation
  using leave-one-out cross-validation and {WAIC},'' \emph{Statistics and
  Computing}, vol.~27, no.~5, pp. 1413--1432, Aug. 2016. [Online]. Available:
  \url{https://doi.org/10.1007/s11222-016-9696-4}
\BIBentrySTDinterwordspacing

\bibitem{volkel2019understanding}
\BIBentryALTinterwordspacing
S.~T. Völkel, D.~Buschek, J.~Pranjic, and H.~Hussmann, ``Understanding emoji
  interpretation through user personality and message context,'' in
  \emph{Proceedings of the 21st International Conference on Human-Computer
  Interaction with Mobile Devices and Services}.\hskip 1em plus 0.5em minus
  0.4em\relax ACM, 10 2019, pp. 1--12. [Online]. Available:
  \url{https://dl.acm.org/doi/10.1145/3338286.3340114}
\BIBentrySTDinterwordspacing

\bibitem{wessel2022bots}
\BIBentryALTinterwordspacing
M.~Wessel, A.~Abdellatif, I.~Wiese, T.~Conte, E.~Shihab, M.~A. Gerosa, and
  I.~Steinmacher, ``Bots for pull requests: The good, the bad, and the
  promising,'' in \emph{Proceedings of the 44th International Conference on
  Software Engineering}.\hskip 1em plus 0.5em minus 0.4em\relax ACM, 5 2022,
  pp. 274--286. [Online]. Available:
  \url{https://dl.acm.org/doi/10.1145/3510003.3512765}
\BIBentrySTDinterwordspacing

\bibitem{wessel2018power}
M.~Wessel, B.~M. De~Souza, I.~Steinmacher, I.~S. Wiese, I.~Polato, A.~P.
  Chaves, and M.~A. Gerosa, ``The power of bots: Understanding bots in oss
  projects,'' \emph{Proceedings of the ACM on Human-Computer Interaction},
  vol.~2, no. CSCW, pp. 1--19, 2018.

\bibitem{wessel2020expect}
M.~Wessel, A.~Serebrenik, I.~Wiese, I.~Steinmacher, and M.~A. Gerosa, ``What to
  expect from code review bots on github? a survey with oss maintainers,'' in
  \emph{Proceedings of the 34th Brazilian Symposium on Software Engineering},
  2020, pp. 457--462.

\bibitem{wessel2020inconvenient}
\BIBentryALTinterwordspacing
M.~Wessel and I.~Steinmacher, ``The inconvenient side of software bots on pull
  requests,'' in \emph{Proceedings of the {IEEE}/{ACM} 42nd International
  Conference on Software Engineering Workshops}.\hskip 1em plus 0.5em minus
  0.4em\relax {ACM}, jun 2020. [Online]. Available:
  \url{https://doi.org/10.1145%2F3387940.3391504}
\BIBentrySTDinterwordspacing

\bibitem{wessel2021dont}
M.~Wessel, I.~Wiese, I.~Steinmacher, and M.~A. Gerosa, ``Don't disturb me:
  Challenges of interacting with software bots on open source software
  projects,'' \emph{Proceedings of the ACM on Human-Computer Interaction},
  vol.~5, no. CSCW2, pp. 1--21, 2021.

\bibitem{wyrich2021waiting}
M.~Wyrich, R.~Ghit, T.~Haller, and C.~M{\"u}ller, ``Bots don’t mind waiting,
  do they? comparing the interaction with automatically and manually created
  pull requests,'' in \emph{2021 IEEE/ACM Third International Workshop on Bots
  in Software Engineering (BotSE)}.\hskip 1em plus 0.5em minus 0.4em\relax
  IEEE, 2021, pp. 6--10.

\end{thebibliography}

\end{document}